\documentclass[prl,aps,twocolumn,epsfig,showpacs,superscriptaddress]{revtex4}
\usepackage{amsmath}
\usepackage{amssymb}
\usepackage[dvips]{graphicx}
\usepackage{dcolumn}
\usepackage{bm}
\usepackage[colorlinks=false,dvipdfm]{hyperref} 
\usepackage{setspace}
\usepackage{amsfonts}
\usepackage{rotating}
\usepackage{makeidx}
\usepackage{amsthm}

\begin{document}

\title{Compact Bell Inequalities for Multipartite Experiments}

\author{Yu-Chun Wu}
 \email{wuyuchun@ustc.edu.cn}
 \affiliation{Key Laboratory of Quantum Information, University of Science and Technology of China, 230026 Hefei, China}

 \author{Marek \.Zukowski}
 \email{marek.zukowski@univie.ac.at}
 \affiliation{Institute of Theoretical Physics and Astrophysics, University of Gda\'nsk, PL-80-952 Gda\'nsk, Poland}
 \affiliation{Hefei National Laboratory for Physical Sciences at Microscale and Department of Modern Physics, University of Science and Technology of China, 230026 Hefei, China}

\author{Jing-Ling~Chen}
 \email{cqtchenj@nus.edu.sg}
 \affiliation{Theoretical Physics Division, Chern Institute of Mathematics, Nankai University, 300071 Tianjin , China}
 \affiliation{Centre for Quantum Technologies, National University of Singapore, 3 Science Drive 2, Singapore 117543}

\author{Guang-Can Guo}
 \affiliation{Key Laboratory of Quantum Information, University of Science and Technology of China, 230026 Hefei, China}


\begin{abstract}
A method for construction of the multipartite Clauser-Horne-Shimony-Holt (CHSH) type Bell inequalities,  for the case of local binary observables, is presented. The standard CHSH-type Bell inequalities can be obtained as special cases. A unified framework to establish all kinds of CHSH-type Bell inequalities by increasing step by step the number of observers is given. As an application, compact Bell inequalities, for eight observers, involving just four correlation functions are proposed. They require much less experimental effort than standard methods and thus is experimentally friendly in multi-photon  experiments.
\end{abstract}


\maketitle

Quantum mechanics is incompatible with local realism \cite{bell}. Bell inequalities reveal this fact. The more Bell inequalities we know, the more we know about the boundaries between Einstein's  local realism and the genuinely non-classical areas of quantum physics, which are potentially useful in quantum information applications. For instance, Bell inequalities have gained a utilitarian power in different quantum information tasks, such as quantum key distribution \cite{ekert}, communication complexity \cite{bruk04} and recently random number generation \cite{pironio}.

For $N$ observers, each choosing between two local dichotomic observables, the complete set of  tight CHSH-type \cite{CHSH}
Bell inequalities has been obtained \cite{werwol,zukbru}. Such inequalities have been pointed out to possess a common structure \cite{ycwu}. However, in the case of more complicated situations (i.e., with more local settings, more parties, or more measurement outcomes), it is still an open task to obtain the complete set of tight Bell inequalities.

Many methods have been put forward to establish Bell inequalities for different situations, such as using the algebraic properties of local observables \cite{mermin,wu}, the stabilizer group of quantum states \cite{stablizer}, and so on. Among these methods, there exists a very important method, which bases on the fact that the set of local realistic models forms a polytope \cite{pit0}, usually called the correlation polytope, whose vertices are the deterministic events. The facets of correlation polytope define tight Bell inequalities. Despite the fact that all vertices of such a correlation polytope are known, it is still difficult to determine all its facets \cite{pit1}. In the simpler case of three qubits, with to numerical methods one may obtain all facets of correlation polytope \cite{sliwa,pit}. The geometrical concept of correlation polytopes is also helpful for building some new Bell inequalities for more complicated cases \cite{ycwu,piro,avis,ycwu1}.

Nevertheless, observations of violations of local realism in the case of multipartite correlations are a challenging task. Generation of multipartite entangled states is a challenge itself  \cite{chen}. Usually, polarization entangled states are generated. The two-photon polarization entangled state (Bell states) can be generated using the technique announced in \cite{zeilinger}. Using methods put forward in \cite{zzw}, the three-photon Greenberger-Horne-Zeilinger (GHZ) state can be generated \cite{zeilinger1}. Pan's group has generated five-photon \cite{pan} and six-photon \cite{pan1} GHZ states. Very recently, an eight-photon GHZ entanglement has been generated \cite{huang,pan8}. However, such experiments require long time to gather data to get the value of a correlation function. Thus, it would be very useful to find handy relevant Bell
inequalities allowing to pinpoint violation of local realism, in a multi-photon experiment, with just few values of correlation functions. Note that this would constitute also a handy, device independent, entanglement witness.

The above is our aim. We present a specific structural feature of the multipartite CHSH-type Bell inequalities,  for the case of local binary observables. Its merits are two-fold: first, one can in principle derive all kinds of arbitrary $N$-partite CHSH-type Bell inequalities, by increasing step by step the number of observers. To see this point more clearly, we shall present some concrete examples to show how one can recover some known Bell inequalities. Second, basing on the  structural feature, compact multipartite Bell inequalities with only four correlation functions are derived.

To begin with,
let us recall some facts about Bell inequalities and the correlation polytope. We shall present this for the case of two parties, Alice and Bob. By $\{\hat{A}_1,\hat{A}_2,\ldots,\hat{A}_M\}$ and $\{\hat{B}_1,\hat{B}_2,\ldots,\hat{B}_N\}$ we denote the  dichotomic local observables that they can choose, and the hidden local realistic measurement outcomes by $a_i$ and $b_j$ for $\hat{A}_i$ and $\hat{B}_j$, respectively. We call such a case an $M\times N$ one. For more parties the generalization is obvious. Usually,  Bell inequalities have the following form: \begin{equation}\label{general form}
\sum_i \alpha_i a_i+\sum_j \beta_j b_j+\sum_{ij}\gamma_{ij}a_ib_j
\leq 1,
\end{equation}
where $\alpha_i,\beta_j,\hbox{ and }\gamma_{ij}$ are real coefficients. There are lower order correlation coefficients for $ a_i$ and $b_j$ and higher correlation coefficients for $a_ib_j$. Inequalities with nonzero lower order correlation coefficients are often called Clauser-Horne (CH) type Bell inequalities \cite{CH}, while inequalities involving only the highest order correlations are usually called CHSH-type Bell inequalities \cite{CHSH}. Homogenization and dehomogenization procedures may establish a connection between CHSH-type and CH-type Bell inequalities \cite{ycwu2}.

There are lots of Bell inequalities, some of them equivalent. Bell inequalities are equivalent if they can be transformed to each other by the following operations:
\begin{itemize}\label{equivalent condition}
\item changing the sign of variables: $X_i\rightarrow -X_i$,
\item permuting the variables: $X_i\rightarrow \sigma(X)_i$,
\item permuting the subscript of the variable: \\ $X_i\rightarrow X_{\sigma(i)}$,
\end{itemize}
where $X$ denotes a local realistic result for any party(e.g. $X_i=a_i$ or $X_i=b_i$).
It is obvious that equivalent inequalities share properties, such as the maximal violation factors, the states violating them, and so on. It is enough to investigate one representative of an equivalence class of Bell inequalities.
For example,  in the $2\times 2$ case, the original  CHSH inequality is the only nontrivial one. In the $3\times3$ case, there is only one nontrivial new CH-type Bell inequality \cite{sliwa}. In the following, we consider equivalence classes of Bell inequalities.

From the point of view of the correlation-polytope method, the CHSH-type Bell inequalities correspond faces of the full correlations polytope. We denote them as $\mathcal {F}_{MN}$ for the
two-party case mentioned above (sometimes we shall use $\mathcal {F}_{MN}$ as the name of the polytope itself, as from mathematical standpoint it is defined by its faces). The inequalities associated with facets (the faces of maximal dimension, that is such that the number of linearly independent vertices on the face equals to the dimension of the polytope) of full correlation polytope are tight CHSH-type Bell inequalities. The vectors representing vertices of a full correlation polytope $\mathcal {F}_{MN}$ are expressible by the following tensor product
\begin{equation}\label{chshvertice}
\vec{a}\otimes\vec{b}=(a_1,a_2,\ldots,a_M)\otimes
(b_1,b_2,\ldots,b_N),
\end{equation}
where $a_i$ and $b_j$ are $\pm 1$. Thus the full  correlation CHSH-type polytope rests in an $MN$-dimensional real space. It has an  inversion symmetry about the origin, no face of the polytope may cross the origin, the number of vertices is $2^{M+N-1}$. Its  faces are defined by a homogeneous linear equation:
\begin{equation}\label{chshfacet}
\sum_{ij}\alpha_{ij}a_ib_j=1,
\end{equation}
where $\alpha_{ij}$ are real coefficients, and for all vertices not at  the face the left-hand side must be strictly smaller than $1$. The associated Bell inequalities are tight CHSH-type ones if $MN$ linearly independent vertices satisfy Eq.~(\ref{chshfacet}).

To illustrate our basic idea and for the sake of simplicity, let us consider a three party case.   For more parties the generalization is straightforward. Assume that the observers can choose between $K,M$ and $N$ binary local measurements, respectively.
Let $\vec{a}=(a_1,a_2,\dots,a_K),\,\vec{b}=(b_1,b_2,\dots,b_M)$ and $\vec{c}=(c_1,c_2,\dots,c_N)$ be the vectors of the measurement outcomes of the three parties respectively. Similar to Eq.~(\ref{chshvertice}), the vertices of correlation polytope $\mathcal {F}_{KMN}$ are $\vec{a}\otimes\vec{b}\otimes\vec{c}$.
The general equation of a face has the  form of
\begin{eqnarray}\label{generalform-chsh}
\mathcal {I}&=&\mathcal {I}(\vec{a},\vec{b},\vec{c})=\sum_{i,j,k}\alpha_{ijk}\;a_ib_jc_k\nonumber\\
&=&c_1f_1+\dots+c_N f_N= 1,
\end{eqnarray}
where $f_i'$s are functions of $a_i,b_j$. The polynomials $\mathcal{I}$ in Eq.~(\ref{generalform-chsh}) can be recast to the following equivalent form as
\begin{eqnarray}
\label{structure}
 \mathcal {I}=\frac{-(N-2)c_1+\sum_{k=1}^{N}c_k}{2}\;
\mathcal {B}_1+\sum_{\ell=2}^{N}\frac{c_1-c_\ell}{2} \;\mathcal
{B}_\ell,
  \end{eqnarray}
with
\begin{eqnarray}
\mathcal {B}_1=\mathcal {I}|_{\vec{c}=\vec{c}_1}=\sum_{k=1}^{N}f_k,
\;\;\;\mathcal {B}_\ell=\mathcal
{I}|_{\vec{c}=\vec{c}_\ell}=\sum_{k=1}^{N}f_k -2f_\ell,
\end{eqnarray}
and $\vec{c}_1=(1,1,1,\ldots,1,1)$, $\vec{c}_2=(1,-1,1,\ldots,1,1)$, $\dots$, $\vec{c}_{N-1}=(1,1,1,\ldots,-1,1)$, $\vec{c}_N=(1,1,1,\ldots,1,-1)$.
Hence if $\mathcal {I} = 1$ is a face of polytope $\mathcal{F}_{KMN}$, namely $\mathcal {I}(\vec{a},\vec{b},\vec{c})\leq 1$ for all vertices $\vec{a}\otimes\vec{b}\otimes\vec{c}$, then for the decomposition in Eq.~(\ref{structure}) one obtains
\begin{eqnarray}
\label{relation}
\mathcal {B}_k(\vec{a},\vec{b})\leq 1, \;\;\;\;k=1,
2, \dots, N,
\end{eqnarray}
and at least one of
$$
f_1\pm f_2\pm \dots\pm f_N
$$
is a face of $\mathcal{F}_{KM}$,  because for some vertices one has $\mathcal {I} = 1$.

Eq.~(\ref{structure}) reveals an important structural feature of CHSH-type Bell inequalities for different number of parties. It evidently links two-party CHSH-type Bell inequalities for $\mathcal {F}_{KM}$ with three-party ones linked with polytope
 $\mathcal {F}_{KMN}$. Thus, one  can devise a constructive method to extend  CHSH-type Bell inequalities into new ones with one additional party, which can choose between $N$ binary local observables.

To extend the CHSH-type Bell inequalities from two-party (or $(n-1)$-party) to three-party (or $n$-party), one may perform the following procedure.
One chooses $N$ faces $\mathcal{B}_k(\vec{a},\vec{b})$ (with $k=1,\dots, N$) from the correlation polytope $\mathcal {F}_{KM}$. By definition they satisfy (\ref{relation}). One then substitutes these $\mathcal {B}_k$'s into Eq.~(\ref{structure}), and if additionally one has $\mathcal {I}\leq 1$ for all vertices $\vec{a}\otimes\vec{b}\otimes\vec{c}$, then $\mathcal {I} = 1$ is a face of $\mathcal {F}_{KMN}$.
Furthermore, if all $\mathcal {B}_k = 1$ are defining facets, namely there are $KM$ linearly independent vertices $\{\vec{a}_{ki}\otimes\vec{b}_{ki}|i=1,\dots,KM\}$ such that $\mathcal{B}_k=1$, and $\mathcal {I} \leq 1$ for all vertices $\vec{a}\otimes\vec{b}\otimes\vec{c}$, then we have $KMN$ linearly independent vertices
$\{\vec{a}_{ki}\otimes\vec{b}_{ki}\otimes\vec{c}_k\}$ at the face defined by $\mathcal {I}=1$, hence it is a facet.

Let us present some examples.

\emph{Example 1. } Take
\begin{subequations}\label{exam1}
\begin{align}
&&\mathcal {B}_1=\frac12(a_2 b_1 +a_2 b_2+a_3 b_2-a_3 b_1),\\
&&\mathcal {B}_2=\frac12(-a_1 b_2 -a_1 b_3-a_3 b_3+a_3 b_2),\\
&&\mathcal {B}_3=\frac12(a_1 b_2 +a_2 b_2+a_1 b_3-a_2b_3).
\end{align}
\end{subequations}
Obviously, all three $\mathcal {B}_k\leq 1$ are the tight two-qubit
CHSH inequalities. Substituting Eq. (\ref{exam1}) into Eq.
(\ref{structure}), we arrive at the following three-setting
inequality as
\begin{eqnarray}\label{mar}
\mathcal {I}&=& \frac{1}{4}(a_1 b_2 c_2+a_2 b_2 c_1+a_2 b_1 c_2
+a_2b_2 c_2-a_1
b_3 c_3\nonumber\\
&&-a_3 b_1 c_3-a_3 b_3 c_1+a_2 b_1 c_3+a_1 b_3 c_2+a_3
b_2 c_1\nonumber\\
&&-a_3 b_1 c_2-a_1 b_2 c_3-a_2 b_3 c_1+a_2 b_3 c_3+a_3 b_2
c_3\nonumber\\
&&+a_3 b_3 c_2)\leq 1.
\end{eqnarray}
This inequality recovers the inequality found in Ref. \cite{WBZ} (see Eq.~(28) in \cite{WBZ}). The CHSH-type inequality (\ref{mar}) is tight and possesses and an interesting property.
It can detect all entanglement in  all generalized GHZ states, $|\psi\rangle=\cos\theta |000\rangle +\sin\theta|111\rangle,\quad \theta\in(0,\pi/4)$, whereas  MABK inequalities fail to detect entanglement for the region of $\theta\in (0,\pi/12]$.

\emph{Example 2. } For a trivial case of a single party with $K$ observables, the facets of `correlation' polytope $\mathcal {F}_K$ are $|a_i|=1$, and the faces are $\sum_{i=1}^{K} p_i a_i=1$, with $ \sum_{i=1}^K |p_i|=1$.
There are no nontrivial facets. Consider now a  $4\times 4$ case. Let us take 4 trivial faces of $\mathcal {F}_4$ for Bob:
\begin{eqnarray*}
\mathcal {B}_1&=&\frac23 b_2+\frac13 b_3,\quad \mathcal {B}_2=-\frac13 b_1 + \frac13 b_2 - \frac13 b_4,\\
\mathcal {B}_3&=&-\frac13 b_1 + \frac13 b_2 + \frac13 b_4,\quad
\mathcal {B}_4=\frac13 b_2+\frac23 b_3.
\end{eqnarray*}
By substituting them into  Eq.~(\ref{structure}) we get a CHSH-type Bell inequality of the form
\begin{eqnarray}\label{CHSH44}
\mathcal {I}=\frac16(-2 a_1 b_1 + a_2 b_1 + a_3 b_1 + a_1 b_2  &\nonumber\\
+ a_1 b_3 + a_2 b_2 + a_2 b_3 + a_3 b_2 + a_3 b_3 &\nonumber\\
+ a_4 b_2 - a_4 b_3 + a_2 b_4 - a_3 b_4)\leq 1.
\end{eqnarray}
This inequality is a tight CHSH-type Bell inequality, namely the Collins-Gisin inequality $I_{4422}\le 1$ of Ref.~\cite{gisin}. The most interesting feature of this inequality is that there exist two-qubit states which violate it which do not violate the CHSH inequality \cite{gisin}. By the way, inequality~(\ref{CHSH44}) is a homogenization of the CH-type Bell inequality, for the case $3\times 3$, derived by Sliwa in \cite{sliwa}. Thus,  it is possible to obtain  tight CHSH-type Bell inequalities from non-tight ones for less parties.

Most importantly, if one considers an additional party, allowed to choose from  for $N=2$ settings, then $\mathcal {B}_1= 1$ and $\mathcal {B}_2=1$ are defining facets of  a polytope for $k$ parties {\em if and only if}
\begin{equation}\label{CHSHtwo}
\mathcal {I}=\frac{c_1+c_2}{2}\mathcal {B}_1+\frac{c_1-c_2}{2}\mathcal {B}_2= 1
\end{equation}
is a facet of the polytope for $k+1$ parties, \cite{ycwu}. This result completely describes the correspondence between the facets of $\mathcal {F}_{KM\cdots}$ and the facets of $\mathcal{F}_{KM\cdots 2}$.
Actually, Eq.~(\ref{CHSHtwo}) allows us to recover the well-known MABK polynomials.

Generally, an $n$-party MABK polynomial has a form:
\begin{equation}\label{MABK}
\mathcal {I}_n=\frac{X_1^n+X_2^n}2 \mathcal {I}_{n-1}+\frac{X_1^n-X_2^n}2\mathcal{I}\;'_{n-1},
\end{equation}
where $X^n_i$'s are  observables for $n$-th party ($i=1\textrm{ or }2$ ), $\mathcal{I}_{n-1}$ is a MABK polynomials for the other $n-1$ parties, $\mathcal{I}\;'_{n-1}$ is obtained from $\mathcal {I}_{n-1}$ by interchanging the indices of 1 and 2 (that is the local observables). In other words, $\mathcal{I}\;'_{n-1}$ and $\mathcal{I}_{n-1}$ are equivalent, as the third  rule of  equivalence applies (see the itemized rules). For example, $\mathcal {I}_2$ and   $\mathcal {I}'_2$ are the usual CHSH polynomials. It should be noted that for $n$ parties, an MABK polynomial contains $2^{n-1}$ terms if $n$ is odd, and $2^n$ terms if $n$ is even.

However, in Eq.~(\ref{CHSHtwo}), it does not matter whether  $\mathcal {B}_1$ and $\mathcal {B}_2$ are equivalent or not. If they are defining facets, then $\mathcal {I}$ is a facet. Hence, one may observe that if $\mathcal {B}_1$ and $\mathcal{B}_2$ in Eq.~(\ref{CHSHtwo}) both have the same four terms but two terms have opposite sign, then the resultant $\mathcal{I}$ has only four terms as well. Using this property, we can obtain compact CHSH-type Bell inequalities for multipartite cases involving {\em just four terms}, which is very handy especially for the experimenters as we shall discuss below. All these four-term compact CHSH-type Bell inequalities are tight. The $k$ partite inequalities of such a kind are violated by $k$ partite GHZ states maximally, all by the same factor of 2.

Recently, an eight-photons GHZ entanglement has been observed in experiments \cite{huang,pan8}. In Ref.~\cite{huang},  entanglement of the eight-photon state was detected with an entanglement witnesses. There were no handy CHSH-type Bell inequalities available during the analysis of the experiment. Standard Bell inequalities contain a lot of terms, for example, the eight-party MABK inequality may have up to $256$ terms \cite{mermin} (the universal inequality, encompassing the MABK ones, introduced in \cite{werwol,zukbru} contains $2^8=256$ terms). This must be contrasted with the fact that according to Huang \emph{et al.} \cite{huang}, it took about 70 hours to obtain an expectation value of one term (the same holds for the universal inequality, encompassing the MABK ones, introduced in \cite{werwol,zukbru}). Therefore, an entanglement witness with 9 terms was adopted in \cite{huang}.
 However, an eight-party compact tight  CHSH-type Bell inequality, obtainable with the methods described above
\begin{eqnarray}\label{eight}
\mathcal {I}=\frac{1}{2} (a_1 b_1 c_1 d_1 e_1 f_1 g_1 h_1 - a_2 b_2 c_1 d_2 e_1 f_1 g_1 h_1 +\nonumber\\
  a_2 b_2 c_2 d_1 e_2 f_2 g_2 h_2 + a_1 b_1 c_2 d_2 e_2 f_2 g_2,
h_2)\leq 1,
\end{eqnarray}
would have been much handier.

Effectiveness of entanglement indicators can be estimated by a quantity that we call critical visibility, $v_{\rm crit}$. It is such a value of $v$ in the mixture
\begin{equation}\label{mixture}
\rho=v|\Psi\rangle\langle\Psi|+(1-v)\rho_{\rm noise}
\end{equation}
for which the given Bell inequality (or entanglement witness) stops to detect the entanglement present in the state $|\Psi\rangle$. Here $\rho_{\rm noise}$ represents the white noise state, which is a unit operator divided by the dimension of the Hilbert space for the considered system. The lower is the value $v_{\rm crit}$ the more robust is the given entanglement indicator. The inequality~(\ref{eight}) has $4$ terms and the required visibility is 0.5, thus its violation can be more robust than in the case of the $9$-term entanglement witness used in the experiment of \cite{huang}, for which the critical visibility   is close to $0.6$. Note that violations of Bell inequalities indicate reductions of communication complexity in various protocols \cite{bruk04}, while entanglement witnesses, basically, just detect entanglement. In this context please note that it is well-known that in general an entangled mixed state may admit a local realistic model in some experimental situations \cite{WERNER}.

In summary, we have pinpointed a structural feature of  multipartite CHSH-type Bell inequalities, for  local binary observables.
It allows  one,  in principle, to derive all kinds of arbitrary $N$-partite CHSH-type Bell inequalities, by increasing step by step the number of observers. Furthermore, using the dehomogenization technique in \cite{ycwu2}, it is possible to obtain the CH-type Bell inequalities. The method presented here allows one to derive multi-party inequalities involving just four terms. This reduces significantly the potential experimental effort required to detect entanglement in multi qubit-experiments. Recently, the creation of GHZ states for 14 qubits has been reported in trapped-ion systems \cite{ino}. The approach presented here could be immensely useful in analysis of  experiments of such a kind.

Y.-C.Wu acknowledges discussions with Prof. C.-F. Li. The work was supported by the National Basic Research Program of China (Grants No.2011CBA00200, 2011CB921200, and 2012CB921900) and National Natural Science Foundation of China (Grant No.10974193, 11275182, 60921091, 10975075, 11175089). M.Z. acknowledges support of MNiSW Grant IdP2011 000361 (Ideas Plus), and a CAS visiting professorship at USTC. The work is partially supported by the National Research Foundation and the Ministry of Education of Singapore.

\end{document}